\documentclass[runningheads]{llncs}
\usepackage{graphicx}
\usepackage{placeins}
\usepackage{color}
\usepackage{stix}
\usepackage{multirow}
\usepackage{xcolor}
\usepackage{color,soul}

\usepackage{ifthen}
\newboolean{anonymous}  
\setboolean{anonymous}{false} 

\usepackage{rotating}

\begin{document}
\title{SER\_AMPEL: a multi-source dataset for speech emotion recognition of Italian older adults \thanks{This research is partially supported by the FONDAZIONE CARIPLO “AMPEL: Artificial intelligence facing Multidimensional Poverty in ELderly” (Ref. 2020-0232) and by the co-funding European Union – Next Generation EU, in the context of The National Recovery and Resilience Plan, Investment Partenariato Esteso PE8 "Conseguenze e sfide dell'invecchiamento", Project Age-It (Ageing Well in an Ageing Society)}}
\titlerunning{SER\_AMPEL: A multi-source dataset for SER of Italian older adults}
%
\ifthenelse{\boolean{anonymous}}{
\author{Anonymous Author(s)*}}
{
\author{Alessandra Grossi\inst{1,2}\orcidID{0000-0003-1308-8497} \and
Francesca Gasparini\inst{1,2}\orcidID{0000-0002-6279-6660} }}
\ifthenelse{\boolean{anonymous}}{}{
\authorrunning{A.Grossi et al.}}

%
\ifthenelse{\boolean{anonymous}}{\institute{}}{
\institute{Department of Informatics, Systems and Communication,\\ University of Milano-Bicocca, Viale Sarca 336, 20126 Milano, Italy\\ 
\url{https://mmsp.unimib.it/}\\
\email{\{alessandra.grossi,francesca.gasparini\}@unimib.it}
\and
NeuroMI, Milan Center for Neuroscience, \\Piazza dell'Ateneo Nuovo 1, 20126 Milano, Italy
}}
\maketitle              
\begin{abstract}
In this paper, SER\_AMPEL, a multi-source dataset for speech emotion recognition (SER) is presented. The peculiarity of the dataset is that it is collected with the aim of providing a reference for speech emotion recognition in case of Italian older adults. The dataset is collected following different protocols, in particular considering acted conversations, extracted from movies and TV series, and recording natural conversations where the
emotions are elicited by proper questions. The evidence of the need for such a dataset emerges from the analysis of the state of the art. Preliminary considerations on the critical issues of SER are reported analyzing  the classification results on a subset of the proposed dataset.

\end{abstract}
\section{Introduction}
\label{sec:intro}
In recent decades, the elderly population has undergone continuous growth with a consequent increase in attention to well-being and active ageing. In this context, loneliness and social isolation of the elderly are factors to consider because they have a close relationship with depression, dementia and other serious medical conditions \cite{national2020social}.
The use of technology can reduce loneliness and social isolation, mitigating their negative outcomes on mental and physical health. In particular, there is a grown interest in studying social robots to increase psychological well being \cite{fraune2022socially}  \cite{chen2018social}. 
As well as in the case of human-human interactions 
\cite{lange2022reading}, emotions play a relevant role in creating an effective interaction between older adults and social robots.  
During social interactions, human beings convey their emotions not only by words but also by bodily, vocal or facial expressions \cite{lange2022reading}. 

In this work we focus on Speech Emotion Recognition (SER) \cite{wu2022design} \cite{akccay2020speech} to recognize emotions in natural conversations, to provide a more natural and social communication in human-robot interactions \cite{hegel2006playing} \cite{jones2008affective}. 

Several researches have been developed in the field of SER during the last three decades \cite{fahad2021survey}, starting from uni-modal approaches, that consider only linguistic or acoustic information, to multi-modal approaches that rely on both of them \cite{atmaja2022survey}.
Several traditional approaches that adopt handcrafted features, both in the temporal and/or frequency domains as well as deep learning ones, have been adopted and are summarized in review manuscripts such as the ones of Akçay et al \cite{akccay2020speech} and Wani et al \cite{wani2021comprehensive}.

Despite the numerous works in the state of the art that face  the SER task, there are several critical issues that make it difficult to recognize emotions especially in natural conversations. In the review paper by Fahad et al. \cite{fahad2021survey}  some of these challenges and the approaches developed to address them are summarized. In particular, SER models struggle to recognize emotions when considering people of different languages or ages.
Moreover, most of the datasets available in the literature are composed of acted utterances \cite{ringeval2013introducing}, while only few of them consider natural conversations \cite{steidl2009automatic} \cite{fan2021lssed} \cite{morrison2007ensemble}. 
\newline
Within this context, we address the problem of SER in the case of an elderly population that speaks Italian.
To this end, the first urgency is the collection of a proper dataset of Italian conversations that considers different ages, in particular older people.

This paper is thus structured as follows. In section \ref{sec:dataset}, the datasets available in the literature are introduced, analyzing their limits. In section \ref{sec:SER_AMPEL}, our multi-source  SER\_AMPEL dataset, collected to solve some of the critical issues found in the available datasets, is presented. Models previously trained on acted datasets are then applied on a portion of the SER\_AMPEL data, and the obtained classification results are reported in section \ref{sec:results}. From this analysis, interesting considerations can be drawn and future lines of research can be identified that are discussed in section \ref{sec:conclusion}.     

\section{SER datasets}
\label{sec:dataset}
Several datasets have been considered in the literature for SER purpose. Following the suggestions proposed by Koolagudi et al. \cite{koolagudi2012emotion}, they can be classified into three groups, depending on how the emotions are elicited. 

\begin{itemize}
    \item \textbf{Acted datasets}: these datasets collect utterances from actors/actresses that try to simulate emotions; the utterances can be short phrases (few seconds), usually repeated with different emotions. The utterances can also be phrases without meaning. Most of the datasets available belong to this category \cite{burkhardt2005database}\cite{cao2014crema}\cite{busso2008iemocap}\cite{livingstone2018ryerson}
    \item \textbf{Evoked or Elicited datasets}: conversations are recorded in situations properly created in order to evoke certain emotions \cite{martin2006enterface}\cite{fan2021lssed}.
    \item \textbf{Spontaneous or Natural datasets}: conversations among different persons are recorded in real-world environments, such as call-centers, or public places \cite{fahad2021survey}\cite{batliner2008releasing}\cite{schuller2020interspeech}. These datasets  are more realistic and authentic, but collecting balanced instances of each emotion is really difficult.
\end{itemize} 

\noindent
We have analyzed 46 datasets described in the literature and reported our analysis in a table available at the following link: 
\ifthenelse{\boolean{anonymous}}{\textit{(Link not available to keep authors anonymous. The related document has been sent with the manuscript as supplementary material)}}{
\url{https://mmsp.unimib.it/download-1/}}.
In this table, among the various information, we have reported the name and reference of the dataset, if it is freely available, number of subjects, gender and age, language, type of collection process (with respect to the three categories introduced), and in particular if there are older adults among the subjects. 

\begin{figure}[!tb]
   \centering
       \includegraphics[width=1\linewidth]{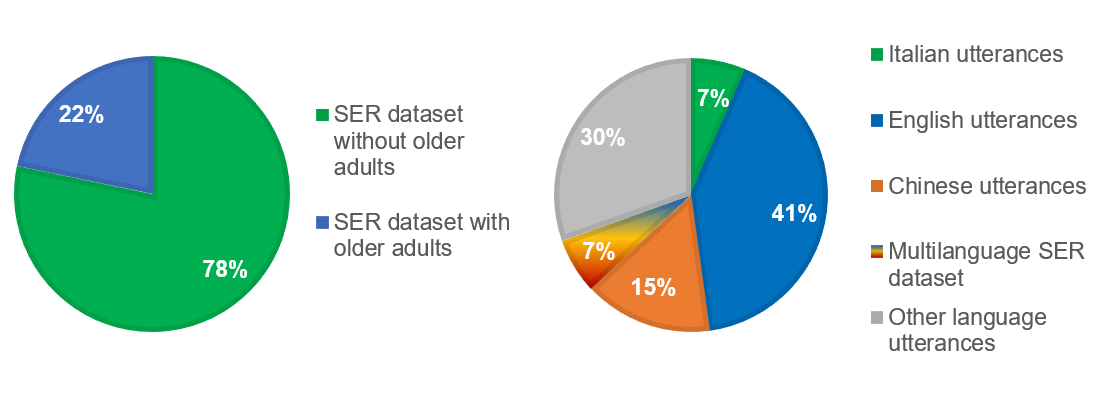}
   \caption{Pie charts depicting the SER datasets that include older subjects with respect to all the others (left) and depicting the relative proportions among datasets of different languages, right.}
   \label{fig:pie}
\end{figure}

From this analysis it emerges that: i) the languages of the collected datasets are mainly English and Chinese, as depicted in the pie chart of 
Figure \ref{fig:pie}, right; ii) multi-language datasets have been proposed \cite{parada2018categorical} \cite{hozjan2002interface}  as language could have an influence in how emotions are expressed \cite{latif2018cross}; iii) most of the available dataset are acted;
iv)  few works face the problem of SER in case of elderly, or varying the age \cite{boateng2020speech} \cite{jian2022speech} \cite{verma2016age} \cite{souganciouglu2020everything}, and old subjects are rarely present in available datasets \cite{schuller2020interspeech} \cite{cao2014crema} \cite{SP2/E8H2MF_2020} \cite{fan2021lssed}, as also depicted in the pie chart of 
Figure \ref{fig:pie}, left; v) the number of available datasets is 12 out of 46.


\section{SER\_AMPEL: a multi-source dataset}
\label{sec:SER_AMPEL}

The SER\_AMPEL is a multi-source dataset  
that has the  intent of providing a reference in case of SER for Italian older adults. It is composed of different subsets, acquired following different protocols. 

The three subsets of the SER\_AMPEL dataset are:   
\begin{itemize}
    \item NOLD \textit{evoked subset}: this is a dataset of natural conversations among Auser \footnote{Auser is an Italian association that promotes active ageing and that supports older persons, www.auser.it} volunteers and older adults, where the emotions are elicited by proper questions and/or music. The list of all the 43 questions is reported in appendix \ref{sec:appendix}. Note that for each conversations only few questions, between 2 and 15 depending on the protocol, were used. Instead, the audio files adopted were chosen by the subject himself/herself. 
    \item NYNG \textit{evoked subset}: this is a dataset of natural conversations among young adults where the emotions are elicited in the same way of NOLD.
    \item AOLD \textit{acted subset}: portion of conversations of Italian older dubbers segmented from Series and Movies. 
\end{itemize}

The SER\_AMPEL dataset is not yet completed, as the acquisition of some subsets is still ongoing, as well as the segmentation, labelling and transcription steps.\\ In particular, the required steps are: 
\begin{enumerate}
    \item segment all the recorded conversations into phrases of about 15 seconds;
    \item label all the segments with respect to two different emotional models: the continuous  Valence-Arousal-Dominance model, introduced by Mehrabian and Russell \cite{russell1977evidence} and the categorical one defined by Ekman \cite{ekman1992there};
    \item transcribe the speech into text, automatically and/or manually.
\end{enumerate}

 In Table \ref{tab:serampel} a synthetic description of the subsets of data already collected is reported and described, considering the classification introduced in section \ref{sec:dataset}, the number of subjects and gender, and type of transcription.

\begin{table}[htbp]
  \centering
  \caption{Main characteristics of each subset of SER\_AMPEL dataset.}
  \renewcommand{\arraystretch}{1.2}
    \begin{tabular}{|c|ccc|ccc|}
    \hline
    \multicolumn{1}{|p{4em}|}{\centering \textbf{Subset}} & \multicolumn{1}{p{4.665em}}{\centering \textbf{No. of \newline{} Speakers}} & 
    \multicolumn{1}{p{4.5em}}{\centering \textbf{No. of \newline{} Female}} & 
    \multicolumn{1}{p{6.5em}|}{\centering \textbf{Age (in years): \newline{}$\mu \pm \sigma$}} & \multicolumn{1}{p{6em}}{\centering  \textbf{No of \newline{} Recordings }} & 
    \multicolumn{1}{p{5em}}{\centering \textbf{Average duration}} & \multicolumn{1}{p{7.3em}|}{\centering  \textbf{Transcriptions}} \\
    \hline
    NOLD & 81 & 31 & 73.03 ± 5.4 & 591 & 40 sec & manual \\
    NYNG & 10 & 6  & 31.09 ± 11,38 & 150 & 1 min & manual \\
    AOLD & 10 & 3  & 71.4 ± 5.3 & 120 & 15 sec & automatic \\
    \hline
    \end{tabular}%
    \renewcommand{\arraystretch}{1}
  \label{tab:serampel}%
\end{table}%

\section{Method and Results}
\label{sec:results}


A general SER model based on XGBoost \cite{chen2016xgboost}, that considers only the acoustic part of the speech, was originally trained on a dataset of English utterances, acted by young and adult actors and actresses \cite{cao2014crema}. 
The classifier has been defined in order to recognize the three sentiment classes: positive, neutral and negative. 
Starting from this model, two domain adaptation strategies (Kullback-Leiber Important Estimation Procedure (KLIEP) \cite{sugiyama2007direct} and Transfer AdaBoost (TrAdaBoost) \cite{inproceedings}) have been proposed to adapt the model to older people and to Italian language respectively \cite{gasparini2022sentiment}. 
The dataset adopted to train the general model and the two adapted ones are CREMA-D \cite{cao2014crema} and EMOVO \cite{costantini2014emovo}. To better understand what follows, a brief description of these two datasets is necessary. 

CREMA-D is an English acted dataset, that consists of 7442 audio and video recordings of professional actors, playing 12 utterances each one expressed in six emotional states (happy, sad, anger, fear, disgust and neutral) at different intensity levels. The utterances last about 5 seconds. All the acted sentences have a neutral semantic content. 91 actors and actresses have been involved (43 female), among them 6 are older persons ($>=$ 60 years) and 85 adults (between 20 and 59 years). Data collected from adults are used to train the general model, while data coming from the older persons (hereafter called OLD) are used to adapt it. 

EMOVO is an Italian acted dataset of 588 audio recordings. Six young Italian actors (3 males and 3 females) with a mean age of 27.1 years, acted 14 utterances, simulating 7 emotional states: neutral, disgust, fear, anger, joy, surprise, and sadness. This dataset (hereafter called ITA) is used to adapt the general model for Italian speakers. 

The best results obtained in the previous paper are reported in Table \ref{tab:prevres}. The models, without domain adaptation (DA) and with DA, trained and tested on the same English dataset (CREMA-D) perform better than the models where training and test sets are of different languages. The domain adaptation strategy applied to adapt the general model (trained on English utterances) to the Italian dataset (ITA test set), increases the performance with respect to the general model without DA. While, in the case of the OLD test set, it seems that there are no differences in the performance of the general model and the DA one.

\begin{table}[h!]
  \centering
  \caption{Comparison of the models using, as test set, the older persons of the CREMA-D dataset (OLD), and the EMOVO dataset (ITA) respectively. The results are reported considering the general model without domain adaptation (No DA) and the best DA strategy (TrAdaBoost). Three evaluation metrics are considered: Macro F1-score, Accuracy and single class F1-score.}
  \hspace*{-0.8cm}
  \renewcommand{\arraystretch}{1.4}
    \begin{tabular}{|c|cc|c|ccc|c|}
    \hline
    \hline
    \textbf{Classifier} & \textbf{DA Strategy} & \multicolumn{1}{p{6em}|}{\centering \textbf{Test Set}} & \multicolumn{1}{p{4em}|}{\centering \textbf{Macro F1-score}} & \multicolumn{1}{p{4.22em}}{\textbf{Negative F1-score}} & \multicolumn{1}{p{3.945em}}{\textbf{Neutral F1-score}} & \multicolumn{1}{p{3.945em}|}{\textbf{Positive F1-score}} & \textbf{Accuracy} \\
    \hline
    \multirow{3}[4]{*}{\begin{sideways}XGBoost \end{sideways}} 
    & \multirow{2}[2]{*}{No DA } & OLD &  \normalsize 62\% &  \normalsize 0,79  & \normalsize 0,55  &  0,51  &  \normalsize 70\% \\
    &    & ITA &  \normalsize 35\% & \normalsize 0,71 &  \normalsize 0,28  &  0,07 &  \normalsize 57\% \\
    \cline{2-8}       
    & \multirow{2}[2]{*}{TrAdaBoost} & OLD & \normalsize 62\% & \normalsize 0,79 & \normalsize 0,56 & \normalsize 0,52 & \normalsize 70\% \\
    &    & ITA & \normalsize 44\% & \normalsize 0,68 & \normalsize 0,25 & \normalsize 0,39 & \normalsize 56\%  \\
   \hline
    \hline
    \end{tabular}%
  \renewcommand{\arraystretch}{1}
  \label{tab:prevres}%
\end{table}%

In Table \ref{tab:performance_DA_ita}, the results obtained applying these models to AOLD subset are reported. 
AOLD is adopted as test set as it is already labelled and transcribed, while the rest of the SER\_AMPEL dataset still requires either labelling and/or transcription. 

In order to use the pre-trained models on this new dataset, the instances of AOLD  have been labelled as positive, neutral and negative. The values of valence have been used to define these classes, considering the high valence scores (5-4) as positive sentiment, low valence scores (1-2) as negative sentiment, and the single central valence score (3) as neutral sentiment. Due to the fact that the dataset collection is still on going, the cardinality of the three groups is unbalanced, with 89 instances labeled as negative, 27 as positive and only 3 neutral. 
The performance while using the general model trained on English adults, is aligned with the results achieved with the same model on the ITA dataset. 
These low performances can be attributed both to the different language between training and test sets, but also to the  presence of different utterances. Note that, the higher result obtained in Table \ref{tab:prevres} first row, is referred to a model where training and test set come from the same dataset, where the utterances pronounced by adults (training) and older people (test) are the same. 
Moreover the DA strategies, seem to be not effective, both in the case of trying to adapt to Italian language and to age.
A difference between CREMA-D and EMOVO datasets and the AOLD one is related to the phrases pronounced. In fact the AOLD dataset is composed of realistic conversations of about 15 seconds, while the other two datasets contain only short utterances, of about 5 seconds, that in some cases correspond to no-sense phrases.

\begin{table}[htbp]
  \centering
  \caption{Performance comparison using the young and adults data from CREMA-D as training set and AOLD data as test set. Two different datasets were adopted as target set for the domain adaptation: Italian acted utterances from EMOVO (rows 2-3) and older adults data from CREMA-D. The analysis are performed considering three different Domain Adaptation (DA) strategies (no Domain Adaptation, KLIEP DA and TrAdaBoost DA). Three evaluation metrics are considered: macro F1-score, accuracy, and single class F1-score.}
  \renewcommand{\arraystretch}{1.4}
    \begin{tabular}{|c|c|ccc|c|}
    \hline
    \hline
    & \multicolumn{1}{p{5.5em}|}{\centering\textbf{Macro F1-score}} 
    & \multicolumn{1}{p{5em}}{\centering\textbf{Negative F1-score}} 
    & \multicolumn{1}{p{5.5em}}{\centering\textbf{Neutral F1-score}} 
    & \multicolumn{1}{p{5.5em}|}{\centering\textbf{Positive F1-score}} & \multicolumn{1}{p{5.5em}|}{\centering\textbf{Accuracy}} \\
    \hline
    \multicolumn{1}{|p{10.5em}|}{\centering No Domain Adaptation} & 32\% & 0,74 & 0,00 & 0,22 & 60\% \\
    \hline
    KLIEP (DA on Ita) & 36\% & 0,85 & 0,00 & 0,35 & 75\% \\
    TrAdaBoost (DA on Ita)& 30\% & 0,42 & 0,14 & 0,35 & 37\% \\
    \hline
    KLIEP (DA on Eld) & 36\% & 0,67 & 0,00 & 0,40 & 56\% \\
    TrAdaBoost (DA on Eld) & 31\% & 0,76 & 0,00 & 0,17 & 62\% \\
    \hline
    \hline
    \end{tabular}%
  \label{tab:performance_DA_ita}%
  \renewcommand{\arraystretch}{1}
\end{table}%

\section{Conclusions}
\label{sec:conclusion}
The development of SER algorithms that adapt their behaviour to different conditions and scenarios is a fundamental step for the definition of systems able to interact in a natural and intuitive way with fragile people such as older adults. These technologies could be employed, for instance, in call centers or mobile communications for the definition of services that adapt their behaviour according to the interlocutor's emotions and mood \cite{gupta2007two}. Furthermore, similar systems could be embedded into nursing-care or assistive robots that continuously interact with older adults or patients to help them in their daily lives. From the subject's voice, the robot could be able to infer the emotional state of the speaker and adapt its behaviour in order to define a more efficient and trustworthy  interaction between robot and people \cite{spezialetti2020emotion}.
In this paper the SER task is specifically considered with respect to Italian older people. From the results here obtained, it emerges the need of a specific dataset, uniform both for what concerns the language and the age of the subjects. To guarantee the generalizability of the SER models that can be developed, multi sources for data collection should be considered. The SER\_AMPEL dataset, here introduced and still ongoing, is conceived with these aims.
Such a dataset could ensure a large collection of speech data useful for the definition of Italian SER algorithms that take into account several factors, for example, the subject's age, the mode of acquisition, or how the emotions are elicited.  Furthermore, the dataset could allow the analysis of issues that may affect speech data collected in natural environments, such as the presence of different kinds of noise as well as the occurrence of utterances or words in dialect.






%
%

\bibliographystyle{splncs04}
\bibliography{mybibliography}

\begin{thebibliography}{10}
\providecommand{\url}[1]{\texttt{#1}}
\providecommand{\urlprefix}{URL }
\providecommand{\doi}[1]{https://doi.org/#1}

\bibitem{akccay2020speech}
Ak{\c{c}}ay, M.B., O{\u{g}}uz, K.: Speech emotion recognition: Emotional
  models, databases, features, preprocessing methods, supporting modalities,
  and classifiers. Speech Communication  \textbf{116},  56--76 (2020)

\bibitem{atmaja2022survey}
Atmaja, B.T., Sasou, A., Akagi, M.: Survey on bimodal speech emotion
  recognition from acoustic and linguistic information fusion. Speech
  Communication  (2022)

\bibitem{batliner2008releasing}
Batliner, A., Steidl, S., N{\"o}th, E.: Releasing a thoroughly annotated and
  processed spontaneous emotional database: the fau aibo emotion corpus  (2008)

\bibitem{boateng2020speech}
Boateng, G., Kowatsch, T.: Speech emotion recognition among elderly individuals
  using multimodal fusion and transfer learning. In: Companion Publication of
  the 2020 International Conference on Multimodal Interaction. pp. 12--16
  (2020)

\bibitem{burkhardt2005database}
Burkhardt, F., Paeschke, A., Rolfes, M., Sendlmeier, W.F., Weiss, B., et~al.: A
  database of german emotional speech. In: Interspeech. vol.~5, pp. 1517--1520
  (2005)

\bibitem{busso2008iemocap}
Busso, C., Bulut, M., Lee, C.C., Kazemzadeh, A., Mower, E., Kim, S., Chang,
  J.N., Lee, S., Narayanan, S.S.: Iemocap: Interactive emotional dyadic motion
  capture database. Language resources and evaluation  \textbf{42},  335--359
  (2008)

\bibitem{cao2014crema}
Cao, H., Cooper, D.G., Keutmann, M.K., Gur, R.C., Nenkova, A., Verma, R.:
  Crema-d: Crowd-sourced emotional multimodal actors dataset. IEEE transactions
  on affective computing  \textbf{5}(4),  377--390 (2014)

\bibitem{chen2018social}
Chen, S.C., Jones, C., Moyle, W.: Social robots for depression in older adults:
  a systematic review. Journal of Nursing Scholarship  \textbf{50}(6),
  612--622 (2018)

\bibitem{chen2016xgboost}
Chen, T., Guestrin, C.: Xgboost: A scalable tree boosting system. In:
  Proceedings of the 22nd acm sigkdd international conference on knowledge
  discovery and data mining. pp. 785--794 (2016)

\bibitem{costantini2014emovo}
Costantini, G., Iaderola, I., Paoloni, A., Todisco, M.: Emovo corpus: an
  italian emotional speech database. In: International Conference on Language
  Resources and Evaluation (LREC 2014). pp. 3501--3504. European Language
  Resources Association (ELRA) (2014)

\bibitem{inproceedings}
Dai, W., Yang, Q., Xue, G.R., Yu, Y.: Boosting for transfer learning. vol.~227,
  pp. 193--200 (01 2007). \doi{10.1145/1273496.1273521}

\bibitem{ekman1992there}
Ekman, P.: Are there basic emotions?  (1992)

\bibitem{fahad2021survey}
Fahad, M.S., Ranjan, A., Yadav, J., Deepak, A.: A survey of speech emotion
  recognition in natural environment. Digital Signal Processing  \textbf{110},
  102951 (2021)

\bibitem{fan2021lssed}
Fan, W., Xu, X., Xing, X., Chen, W., Huang, D.: Lssed: a large-scale dataset
  and benchmark for speech emotion recognition. In: ICASSP 2021-2021 IEEE
  International Conference on Acoustics, Speech and Signal Processing (ICASSP).
  pp. 641--645. IEEE (2021)

\bibitem{fraune2022socially}
Fraune, M.R., Komatsu, T., Preusse, H.R., Langlois, D.K., Au, R.H., Ling, K.,
  Suda, S., Nakamura, K., Tsui, K.M.: Socially facilitative robots for older
  adults to alleviate social isolation: A participatory design workshop
  approach in the us and japan. Frontiers in Psychology p.~6256 (2022)

\bibitem{gasparini2022sentiment}
Gasparini, F., Grossi, A.: Sentiment recognition of italian elderly through
  domain adaptation on cross-corpus speech dataset. In: CEUR Workshop
  Proceeding. vol.~3367, pp. 12--28 (2022)

\bibitem{gupta2007two}
Gupta, P., Rajput, N.: Two-stream emotion recognition for call center
  monitoring. In: Eighth Annual Conference of the International Speech
  Communication Association. Citeseer (2007)

\bibitem{hegel2006playing}
Hegel, F., Spexard, T., Wrede, B., Horstmann, G., Vogt, T.: Playing a different
  imitation game: Interaction with an empathic android robot. In: 2006 6th
  IEEE-RAS International Conference on Humanoid Robots. pp. 56--61. IEEE (2006)

\bibitem{hozjan2002interface}
Hozjan, V., Kacic, Z., Moreno, A., Bonafonte, A., Nogueiras, A.: Interface
  databases: Design and collection of a multilingual emotional speech database.
  In: LREC (2002)

\bibitem{jian2022speech}
Jian, Q., Xiang, M., Huang, W.: A speech emotion recognition method for the
  elderly based on feature fusion and attention mechanism. In: Third
  International Conference on Electronics and Communication; Network and
  Computer Technology (ECNCT 2021). vol. 12167, pp. 398--403. SPIE (2022)

\bibitem{jones2008affective}
Jones, C., Deeming, A.: Affective human-robotic interaction. In: Affect and
  emotion in human-computer interaction, pp. 175--185. Springer (2008)

\bibitem{koolagudi2012emotion}
Koolagudi, S.G., Rao, K.S.: Emotion recognition from speech: a review.
  International journal of speech technology  \textbf{15}(2),  99--117 (2012)

\bibitem{lange2022reading}
Lange, J., Heerdink, M.W., Van~Kleef, G.A.: Reading emotions, reading people:
  Emotion perception and inferences drawn from perceived emotions. Current
  Opinion in Psychology  \textbf{43},  85--90 (2022)

\bibitem{latif2018cross}
Latif, S., Qayyum, A., Usman, M., Qadir, J.: Cross lingual speech emotion
  recognition: Urdu vs. western languages. In: 2018 International Conference on
  Frontiers of Information Technology (FIT). pp. 88--93. IEEE (2018)

\bibitem{livingstone2018ryerson}
Livingstone, S.R., Russo, F.A.: The ryerson audio-visual database of emotional
  speech and song (ravdess): A dynamic, multimodal set of facial and vocal
  expressions in north american english. PloS one  \textbf{13}(5),  e0196391
  (2018)

\bibitem{martin2006enterface}
Martin, O., Kotsia, I., Macq, B., Pitas, I.: The enterface'05 audio-visual
  emotion database. In: 22nd international conference on data engineering
  workshops (ICDEW'06). pp.~8--8. IEEE (2006)

\bibitem{morrison2007ensemble}
Morrison, D., Wang, R., De~Silva, L.C.: Ensemble methods for spoken emotion
  recognition in call-centres. Speech communication  \textbf{49}(2),  98--112
  (2007)

\bibitem{parada2018categorical}
Parada-Cabaleiro, E., Costantini, G., Batliner, A., Baird, A., Schuller, B.:
  Categorical vs dimensional perception of italian emotional speech  (2018)

\bibitem{SP2/E8H2MF_2020}
Pichora-Fuller, M.K., Dupuis, K.: {Toronto emotional speech set (TESS)} (2020).
  \doi{10.5683/SP2/E8H2MF}, \url{https://doi.org/10.5683/SP2/E8H2MF}

\bibitem{ringeval2013introducing}
Ringeval, F., Sonderegger, A., Sauer, J., Lalanne, D.: Introducing the recola
  multimodal corpus of remote collaborative and affective interactions. In:
  2013 10th IEEE international conference and workshops on automatic face and
  gesture recognition (FG). pp.~1--8. IEEE (2013)

\bibitem{russell1977evidence}
Russell, J.A., Mehrabian, A.: Evidence for a three-factor theory of emotions.
  Journal of research in Personality  \textbf{11}(3),  273--294 (1977)

\bibitem{schuller2020interspeech}
Schuller, B.W., Batliner, A., Bergler, C., Messner, E.M., Hamilton, A.,
  Amiriparian, S., Baird, A., Rizos, G., Schmitt, M., Stappen, L., et~al.: The
  interspeech 2020 computational paralinguistics challenge: Elderly emotion,
  breathing \& masks  (2020)

\bibitem{national2020social}
National Academies~of Sciences, E., Medicine, et~al.: Social isolation and
  loneliness in older adults: Opportunities for the health care system.
  National Academies Press (2020)

\bibitem{souganciouglu2020everything}
So{\u{g}}anc{\i}o{\u{g}}lu, G., Verkholyak, O., Kaya, H., Fedotov, D.,
  Cad{\'e}e, T., Salah, A.A., Karpov, A.: Is everything fine, grandma? acoustic
  and linguistic modeling for robust elderly speech emotion recognition. arXiv
  preprint arXiv:2009.03432  (2020)

\bibitem{spezialetti2020emotion}
Spezialetti, M., Placidi, G., Rossi, S.: Emotion recognition for human-robot
  interaction: Recent advances and future perspectives. Frontiers in Robotics
  and AI p.~145 (2020)

\bibitem{steidl2009automatic}
Steidl, S.: Automatic classification of emotion related user states in
  spontaneous children's speech. Logos-Verlag Berlin, Germany (2009)

\bibitem{sugiyama2007direct}
Sugiyama, M., Nakajima, S., Kashima, H., Buenau, P., Kawanabe, M.: Direct
  importance estimation with model selection and its application to covariate
  shift adaptation. Advances in neural information processing systems
  \textbf{20} (2007)

\bibitem{verma2016age}
Verma, D., Mukhopadhyay, D.: Age driven automatic speech emotion recognition
  system. In: 2016 International Conference on Computing, Communication and
  Automation (ICCCA). pp. 1005--1010. IEEE (2016)

\bibitem{wani2021comprehensive}
Wani, T.M., Gunawan, T.S., Qadri, S.A.A., Kartiwi, M., Ambikairajah, E.: A
  comprehensive review of speech emotion recognition systems. IEEE Access
  \textbf{9},  47795--47814 (2021)

\bibitem{wu2022design}
Wu, X., Zhang, Q.: Design of aging smart home products based on radial basis
  function speech emotion recognition. Frontiers in Psychology  \textbf{13},
  882709--882709 (2022)

\end{thebibliography}

\newpage
\appendix

\section{Appendix A}
\label{sec:appendix}

\begin{table}[htbp]
  \centering
  \caption{Questions selected to evoke Positive Sentiment}
  \renewcommand{\arraystretch}{1.2}
    \begin{tabular}{|c|l|}
    \hline
    \multicolumn{1}{|p{3.5em}|}{\centering\textbf{ID}} &  \multicolumn{1}{|p{40em}|}{\centering\textbf{Questions}} \\
    \hline
    Q\_01 & Tell us about a fun adventure you had with your friends\\
    Q\_02 & What is your favourite time of day? Please, describe it \\
    Q\_03 & Are there any hobbies or activities you began during the lockdown? If so, please describe them \\
    Q\_04 & Where would you like to go on vacation? \\
    Q\_05 & What is your favorite season and why?\\
    Q\_06 & Tell us about the happiest moment of your life \\
    Q\_07 & Do you have grandchildren? Please, tell us about them\\
    Q\_08 & Describe a happy moment spent with your parents\\
    Q\_09 & Are you married? If so, how did you meet each other? \\
    Q\_10 & Have you ever traveled? If so, tell us about your most beautiful trip \\
    Q\_11 & Describe the happiest moment you experienced during the lockdown \\
    Q\_12 & \multicolumn{1}{p{39em}|}{Were you able to keep in touch with friends and relatives during the lockdown? If so, please describe how} \\
    Q\_13 & Has the lockdown period allowed you to rediscover activities or passions that you had set aside? \\
    Q\_14 & What did you enjoy most during the lockdown period? \\
    \hline
    \end{tabular}%
    \renewcommand{\arraystretch}{1}
  \label{tab:question_pos}%
\end{table}%

\begin{table}[!h]
  \centering
  \caption{Questions selected to evoke Neutral Sentiment}
  \renewcommand{\arraystretch}{1.2}
    \begin{tabular}{|c|l|}
    \hline
    \multicolumn{1}{|p{3.5em}|}{\centering\textbf{ID}} &  \multicolumn{1}{|p{40em}|}{\centering\textbf{Questions}} \\
    \hline
    Q\_15 & Please, describe your ordinary day / Which activities do you daily perform? \\
    Q\_16 & Describe in detail the path  to go shopping \\
    Q\_17 & Where do you usually read the news? \\
    Q\_18 & Please describe a room of your home in detail  \\
    Q\_19 & What did you have for lunch/dinner?  \\
    Q\_20 & Age, gender, place of birth, educational background and work history \\
    Q\_21 & Please, describe the place where you live \\
    Q\_22 & Please, describe the type of dwelling where you live \\
    Q\_23 & Please, describe what a car looks like \\
    Q\_24 & Please, describe what a chair looks like \\
    \hline
    \end{tabular}%
    \renewcommand{\arraystretch}{1}
  \label{tab:question_neu}%
\end{table}%

\begin{table}[htbp]
  \centering
  \caption{Questions selected to evoke Negative Sentiment}
  \renewcommand{\arraystretch}{1.2}
    \begin{tabular}{|c|l|}
    \hline
    \multicolumn{1}{|p{3.5em}|}{\centering\textbf{ID}} &  \multicolumn{1}{|p{40em}|}{\centering\textbf{Questions}} \\
    \hline
    Q\_25 & Tell us about a news story you recently heard that saddened you?\\
    Q\_26 & Is there anything in your neighborhood you want to change? \\
    Q\_27 & What did you miss most during the lockdown period?\\
    Q\_28 & What do you think about the recent price increase? \\
    Q\_29 & What is the most dramatic moment in someone's life? \\
    Q\_30 & What activities or hobbies have you been unable to carry out during the lockdown? \\
    Q\_31 & Have you ever suffered the loss of a loved one?  \\
    Q\_32 & Have you ever been let down by someone you cared about? \\
    Q\_33 & \multicolumn{1}{p{39em}|}{Has the economic situation ever affected your happiness during any moment of your life?}  \\
    Q\_34 & Do you remember a moment of your childhood that saddened you deeply?  \\
    Q\_35 & Have you ever suffered a disappointment of love?  \\
    Q\_36 & What saddened you most during lockdown?  \\
    Q\_37 & What saddened you most  during the pandemic from a global point of view?  \\
    \hline
    \end{tabular}%
    \renewcommand{\arraystretch}{1}
  \label{tab:question_neg}%
\end{table}%

\begin{table}[!htbp]
  \centering
  \caption{Questions not related to a specific sentiment}
  \renewcommand{\arraystretch}{1.2}
    \begin{tabular}{|c|l|}
    \hline
    \multicolumn{1}{|p{3.5em}|}{\centering\textbf{ID}} &  \multicolumn{1}{|p{40em}|}{\centering\textbf{Questions}} \\
    \hline
    Q\_38 & Do you think the experience of the pandemic could have affected you in any way?\\
    Q\_39 & What is an object you particularly care about and what does it mean for you? \\
    Q\_40 & \multicolumn{1}{p{39em}|}{What is your perception of the area or the city where you live? How do they feel about your neighbourhood? Is there anything you like or dislike? How do you move within your neighbourhood, and from your home to the rest of the city?} \\
    Q\_41 & \multicolumn{1}{p{39em}|}{What is your vision of the future city? How do you imagine the neighbourhood where you live, and the city in general, when your grandchildren will be older adults?} \\
    Q\_42 & \multicolumn{1}{p{39em}|}{Think about of a song you like, would you describe an episode of your life related to that song?} \\
    Q\_43 & What do you think about the experiment you have just performed? \\
    \hline
    \end{tabular}%
    \renewcommand{\arraystretch}{1}
  \label{tab:addlabel}%
\end{table}%
\end{document}